\documentclass[twocolumn,secnumarabic,amssymb, amsmath, nofootinbib,tightenlines,
nobibnotes, aps, prl,epsfig]{revtex4}
\usepackage{graphicx}% Include figure files
\usepackage{dcolumn}% Align table columns on decimal point
\usepackage{bm}% bold math
\begin{document}
\preprint{APS/123-QED}
\title{NLO corrections to the hard Pomeron behavior of the charm structure functions $F_{k}^{c}(k=2,L)$ at low-$x$ }% Force line breaks with \\

\author{G.R.Boroun}%
 \email{grboroun@gmail.com; boroun@razi.ac.ir }
\author{B.Rezaei }
\altaffiliation{brezaei@razi.ac.ir}%Lines break automatically or can be forced with \\
\affiliation{ Physics Department, Razi University, Kermanshah
67149, Iran}% \textbackslash\textbackslash
\date{\today}% It is always \today, today,
\begin{abstract}
%%%%%%%%%%%%%%%%%%%%%%%%%%%%%%%%%%%%%%%%%%%%%%%%%%%%%%%
We show that the charm structure functions $F_{k}^{c}$  have a
hard pomeron behavior at low-$x$, as the gluon distribution  is
dominated  by the hard (Lipatov) pomeron at small $x$ and all
$Q^{2}$ values. It is shown that the charm structure functions
obtained, using the gluon distribution functions, are in agreement
with data from HERA in the next-to-leading analysis (NLO). Having
checked that this behavior gives the charm structure functions
exponent independent of $x$. With respect to the hard (Lipatov)
pomeron for the low $x$ gluon distribution, we provide a compact
formula for the ratio $R^{c}=F_{L}^{c}/F_{2}^{c}$ that is
approximately independent of $x$ at low- $x$ and its independent
of the input gluon distribution function at all $Q^{2}$ values.
 \\
%%%%%%%%%%%%%%%%%%%%%%%%%%%%%%%%%%%%%%%%%%%%%%%%%%%%%%%
\end{abstract}
 %PACS, the Physics and Astronomy
                              %Classification Scheme.
 %Use showkeys class option if keyword
\maketitle
%%%%%%%%%%%%%%%%%%%%%%%%%%%%%%%%%%%%%%%%%%%%%%%%%%%%%%%%%%%%%%%%%
\centerline{\textbf{Introduction}}

The small- $x$ region of the deep inelastic scattering (DIS)
offers a unique possibility to explore the Regge limit of
perturbative quantum chromodynamics (PQCD) [1,2]. This theory is
successfully described by the exchange of a particle with
appropriate quantum numbers and the exchange particle is called a
Regge pole. Phenomenologically, the Regge pole approach to deep
inelastic scattering implies that the structure functions are sums
of powers in $x$. As the simplest fit to the small $x$ data
corresponds to
$F_{2}(x,Q^{2})=\sum_{i}A_{i}(Q^{2})x^{-\lambda_{i}}$, where the
singlet part of the structure function is controlled by pomeron
exchange at small $x$. HERA shows, this behavior for the gluon and
sea quark distributions  at small $x$. As, in the limit
$x{\rightarrow}0$, the gluon distribution will become large, so
its contribution to the evolution of the parton distribution
becomes dominant. Therefore the gluon distribution has a rapid
rise behavior at small $x$ as
\begin{equation} xg(x,Q^{2})= A_{g}x^{-\lambda_{g}},
\end{equation}
where $\lambda_{g}$ is the pomeron intercept minus one. This steep
behavior of the gluon generates a similar steep behavior of
$F^{c}_{k}$ at small $x$.  Because the gluon will drive the quark
singlet distribution, so that the charm structure functions
$F^{c}_{k}$ is expected to rise approximately like a power of $x$.
For $Q^{2}{\leq}1\hspace{0.1cm}$GeV$^{2}$, the simplest Regge
phenomenology predicts that the value of
$\lambda_{S}=\alpha_{\mathbb{P}}(0)-1{\simeq}\hspace{0.1cm}0.08$
is consistent with the hadronic Regge theory [3-5] where
$\alpha_{\mathbb{P}}(0)$ is described soft pomeron dominant with
its intercept slightly above unity ($\sim$1.08). Whereas for
$Q^{2}{\geq}1\hspace{0.1cm}$GeV$^{2}$ the slope rises steadily to
reach a value greater than $0.3$ by
$Q^{2}{\approx}100\hspace{0.1cm}$GeV$^{2}$ where hard pomeron is
dominant. This larger value of $\lambda_{S}$ is not so far from
that expected using BFKL [6] ideas. There are some  other authors
[3-5] who extended their Regge model adding a hard Pomeron with
intercept $1.44$, which allows them to describe the low $x$ HERA
data up to $Q^{2}$ values of a few hundred GeV$^{2}$. HERA data
for the charm structure functions $F^{c}_{k}(x,Q^{2}),(k=2,L)$,
over a wide range of $Q^{2}$, require only a hard pomeron as the
coupling of the soft pomeron to charm is apparently very small.
The charm structure function can be described by a single power of
$x$
\begin{equation}
F^{c}_{k}(x,Q^{2})= A^{c}_{k}x^{-\lambda^{c}_{k}},
\end{equation}
where $\lambda^{c}_{k}$ is pomeron exponent and we expect to
determine. The study of production mechanisms of heavy quarks
provides us with new tests of QCD. As in perturbative QCD (pQCD)
physical quantities can be expanded in the strong coupling
constant $\alpha_{s}(\mu^{2})$. When the scale $\mu$ has to be
large, provided we are dealing with so called hard processes. In
the case of heavy quark production, we can have condition that the
heavy quarks produced from the boson- gluon fusion (BGF) according
to Fig.1. That is, in PQCD calculations the production of heavy
quarks at HERA proceeds dominantly via the direct BGF, where the
photon interacts indirectly with a gluon in the proton by the
exchange of a heavy quark pair [7-13]. In the BGF dynamic, the
charm(beauty) quark is treated as a heavy quark and its
contribution is given by fixed- order perturbative theory. With
respect to the recent measurements of HERA, the charm contribution
to the structure function at small $x$ is a large fraction of the
total [14-15], as this value is approximately $30\%$ ($1\%$)
fraction of the total. This behavior is directly related to the
growth of the gluon distribution at small $x$. We know that the
gluons couple only through the strong interaction, consequently
the gluons are not directly probed in DIS. Only contributing
indirectly is via the
 $g{\rightarrow}q\bar{q}$
transition. This involves the computation of the BGF process
$\gamma^{\star}g{\rightarrow}c\bar{c}$. This process can be
created when the squared invariant mass of the hadronic final
state has the condition that $W^{2}{\geq}4m_{c}^{2}$.\\
In this paper, we investigate the hard (Lipatov)Pomeron behavior
of the charm structure functions and present the ratio of the
$R^{c}=\frac{F^{^{c\overline{c}}}_L}{F^{^{c\overline{c}}}_2}$ with
respect to this behavior which its independent of the input gluon
distribution function.\\

\centerline{\textbf{Charm Structure Functions}}

In the low- $x$ range, where only the gluon contribution is
matter, the charm quark contribution
$F_{k}^{c}(x,Q^{2},m^{2}_{c})(k=2,L)$ to the proton structure
function  is given by this form
\begin{eqnarray}
F_{k}^{c}(x,Q^{2},m^{2}_{c})&=&2e_{c}^{2}\frac{\alpha_{s}(\mu^{2})}{2\pi}\int_{1-\frac{1}{a}}^{1-x}dzC_{g,k}^{c}
(1-z,\zeta)\nonumber\\
&& {\times}G(\frac{x}{1-z},\mu^{2}),
\end{eqnarray}
where $a=1+4\zeta(\zeta{\equiv}\frac{m_{c}^{2}}{Q^{2}})$,
$G(x,\mu^{2})$ is the gluon distribution function and $\mu$ is the
mass factorization scale, which has been put equal to the
renormalization scales $\mu^{2}=4m_{c}^{2}$ or
$\mu^{2}=4m_{c}^{2}+Q^{2}$. Here $C^{c}_{g,k}$ is the charm
coefficient functions in LO and NLO analysis as
\begin{eqnarray}
C_{k,g}(z,\zeta)&{\rightarrow}&C^{0}_{k,g}(z,\zeta)+a_{s}(\mu^{2})[C_{k,g}^{1}(z,\zeta)\\\nonumber
&&+\overline{C}_{k,g}^{1}(z,\zeta)ln\frac{\mu^{2}}{m_{c}^{2}}],
\end{eqnarray}
where $a_{s}(\mu^{2})=\frac{\alpha_{s}(\mu^{2})}{4\pi}$ and in the
NLO analysis
\begin{eqnarray}
\alpha_{s}(\mu^{2})=\frac{4{\pi}}{\beta_{0}ln(\mu^{2}/\Lambda^{2})}
-\frac{4\pi\beta_{1}}{\beta_{0}^{3}}\frac{lnln(\mu^{2}/\Lambda^{2})}{ln(\mu^{2}/\Lambda^{2})}
\end{eqnarray}
with $\beta_{0}=11-\frac{2}{3}n_{f},
\beta_{1}=102-\frac{38}{3}n_{f} $ ($n_{f}$ is the number of active
flavours).\\

In the LO analysis, the coefficient functions BGF can be found
[16], as
\begin{eqnarray}
C^{0}_{g,2}(z,\zeta)&=&\frac{1}{2}([z^{2}+(1-z)^{2}+4z\zeta(1-3z)-8{\zeta^{2}}z^{2}]\nonumber\\
&&{\times}ln\frac{1+\beta}{1-\beta}+{\beta}[-1+8z(1-z)\nonumber\\
&&-4z{\zeta}(1-z)]),
\end{eqnarray}
and
\begin{eqnarray}
C^{0}_{g,L}(z,\zeta)=-4z^{2}{\zeta}ln\frac{1+\beta}{1-\beta}+2{\beta}z(1-z),
\end{eqnarray}
where $\beta^{2}=1-\frac{4z\zeta}{1-z}$.\\
 At NLO,
$O(\alpha_{em}\alpha_{s}^{2})$, the contribution of the photon-
gluon component is usually presented in terms of the coefficient
functions $C_{k,g}^{1}, \overline{C}_{k,g}^{1}$. Using the fact
that  the virtual photon- quark(antiquark) fusion subprocess can
be neglected, because their contributions to the heavy-quark
leptoproduction vanish at LO and are small at NLO [6,17].  In a
wide kinematic range, the contributions to the charm structure
functions in NLO are not positive due to mass factorization.
Therefore the charm structure functions are dependence to the
gluonic observable in LO and NLO. The NLO coefficient functions
are only avaliable as computer codes[17,18]. But in the high-
energy regime ($\zeta<<1$) we can used the compact form of these
coefficients according to
the Refs.[18,19].\\

Exploiting the low- $x$ behavior of the gluon distribution
function according to the hard (Lipatov) Pomeron as
\begin{equation}
G(x,\mu^{2}){\rightarrow}x^{-\lambda_{g}}.
\end{equation}
The power of $\lambda_{g}$ is found to be either $\lambda_{g}
{\simeq} 0$ or $\lambda_{g}{\simeq} 0.5$. The first value
corresponds to the soft Pomeron and the second value to the hard
(Lipatov) Pomeron intercept. Based on the hard (Lipatov) pomeron
behavior for the gluon distribution, let us put Eq.(8) in Eq.(3).
After doing the integration over $z$, Eq.(3) can be rewritten as
\begin{eqnarray}
F_{k}^{c}(x,Q^{2},m^{2}_{c})=2e_{c}^{2}\frac{\alpha_{s}(\mu^{2})}{2\pi}I_{k}(x,\mu^{2})\nonumber\\
{\times}G(x,\mu^{2}),
\end{eqnarray}
where
\begin{eqnarray}
I_{k}(x,\mu^{2})=\int_{1-\frac{1}{a}}^{1-x}C_{g,k}^{c}
(1-z,\zeta)(1-z)^{\lambda_{g}}dz,
\end{eqnarray}
 here $C_{g,k}^{c}$ is defined by Eq.4. We observe that this equation (Eq.9) is directly dependence to the gluon
distribution, which is usually taken from the GRV [16], CETQ [21]
, MRST [22] parameterizations or DL [3-5] model. In what follows
we shall use the gluon distribution with an intercept  according
to the hard- pomeron behavior at the DL model.\\
In fact, the gluon
distribution function input $G(x,\mu^{2})$ does cancels in the
ratio of the charm structure functions as we have
\begin{equation}
R^{c}=\frac{I_{L}(x,\mu^{2})}{I_{2}(x,\mu^{2})}.
\end{equation}
 Therefore, this ratio ,which is independent of the gluon distribution function, is very useful for practical
 applications. In this equation we used the solutions of the NLO
 BGF analysis and considered $\lambda_{g}$ as a hard (Lipatov)
 Pomeron.\\
One striking feature of the Pomeron effective exponents to the
behavior of the charm structure functions are the predictions of
the charm exponents  with respect to Eq.9. Namely,  the charm
exponents $\lambda_{k}^{c}$ at small $x$ are not a function of the
variable $x$. Here we show that the charm structure function
exponents exhibit the hard pomeron behavior, where we have taken
into account that this behavior is given by the gluon distribution
exponent. Therefore, the concept of the hard pomeron with an
intercept that is independent of $Q^{2}$ is supported here for the
charm structure functions. As we have
\begin{eqnarray}
\lambda^{c}_{k} {\simeq}
\lambda_{g}+\frac{{\partial}}{{\partial}{\ln}\frac{1}{x}}{\ln}[\int_{1-\frac{1}{a}}^{1-x}C_{g,k}^{c}
(1-z,\zeta)(1-z)^{\lambda_{g}}dz].\nonumber\\
 \end{eqnarray}

In this equation, the second term of the right hand Eq.12 at any
value of $x$ and all $Q^{2}$ values, is very small. We conclude
that a simplified power- like parameterization  with
$\lambda^{c}_{k} {\simeq} \lambda_{g}$ leads to the violation of
unitarity providing the Froisart- Martin [23] bound, at low- $x$
values. We found that our results for the charm structure
functions exponents at the renormalization scales are the same
hard (Lipatov) pomeron intercept in a wide region of $Q^{2}$ and
$x$ values. Consequently, these results show that
$\lambda^{c}_{k}$ determined within the given errors are constant
at small- $x$. Also, it is a straightforward matter to consider of
the logarithmic derivatives of the charm
 structure functions defined by
 $\frac{{\partial}lnF_{k}^{c}(x,Q^{2})}{{\partial}ln\frac{1}{x}}$ [24].
 We note that the charm
structure functions exponents, when starting from the Regge- like
behavior of the charm structure
 functions, can be written by
\begin{equation}
\lambda_{k}^{c}{\simeq}\frac{{\partial}lnF_{k}^{c}(x,Q^{2})}{{\partial}ln\frac{1}{x}}+
ln(x)\frac{{\partial}\lambda_{k}^{c}}{{\partial}ln\frac{1}{x}}.
\end{equation}\\
With respect to the charm structure functions (Eq.9), we can
define the charm intercepts into the logarithmic $x$-derivatives
of the charm structure functions and the derivatives of the charm
intercepts.\\

 \centerline{\textbf{Results and Discussion}}

In this section, we present numerical analysis of the NLO
corrections to the charm structure functions. In our calculations,
we use the DL parametrization of the gluon distribution and also
we set the running coupling constant with $\Lambda=0.224 GeV$, and
for our input parameters we choose $m_{c}=1.5GeV$. The theoretical
uncertainties in our result are according to the renormalization
scales  $\mu^{2}=4m^{2}_{c}$ and
$\mu^{2}=4m^{2}_{c}+Q^{2}$.\\
 Fig.2 shows the quantity $F_{2}^{c}(x,Q^{2})$ as a function of $x$
for $Q^{2}=45 GeV^{2}$. The LO and NLO predictions are given by
the theoretical uncertainty from the renormalization scales,
correspondingly. The GJR parameterization [25] are presented by
solid curve. One can see that NLO corrections to the BGF kernels
are importance, especially for low $x{<}0.01$.  At the same time,
the different between the NLO and LO corrections to the charm
structure function increase as $x$ decreases, when compared to the
GJR parameterization.\\
 In Figs.3 and 4 we
present the behavior of the $F_{k}^{c}(k=2,L)$
 charm structure functions as a function of $x$ that accompanied with the theoretical uncertainly,
 and compared to the ZEUS and H1 data [13]. Also we compared our results for the
charm structure function to the DL model [3-5](Solid curve) and
the color dipole model [26] (Dash curve). We can observe that our
results are comparable with the experimental data and also with
these models. We can also obtain a reasonable description for the
longitudinal charm structure function, when compared results only
with the color dipole model [26]. We observed that this
corresponding is
good  at all $Q^{2}$ values.\\
 As can be seen in all figures, the increase of  our calculations for the
 charm structure functions $F^{c}_{k}(x,Q^{2})$ towards low
 $x$ are consistent with the experimental data.
  This implies that the $x$ dependence of the
 charm structure functions at low $x$ are consistent with a
 power law, $F^{c}_{k}=A_{k}x^{-\lambda^{c}_{k}}$, for fixed $Q^{2}$.
  Having concluded that the data for $F^{c}_{k}$ require a hard
Pomeron component, as tested this behavior with our results. \\

In Fig.5 we present our results for the ratio
$R^{c}=F_{L}^{c}/F_{2}^{c}$
(=$\frac{I_{L}(x,\mu^{2})}{I_{2}(x,\mu^{2})}$) [18,27-28] in charm
leptoproduction. The theoretical uncertainty  in our results
related to the renormalization scale choices. We observe that this
ratio is independent of $x$ for $x~{\leq}~0.01$ in a wide range of
$Q^{2}$. We see that this value is approximately between $0.10$
and $0.24$ in a region of $Q^2$ and this prediction for $R^{c}$ is
close to the results Refs.[18,26-29]. In Fig.6 we present the
ratio $R^{c}$ as function of $Q^{2}$ at $x=0.001$ with respect to
the theoretical uncertainty. We can see that the behavior of this
ratio is agree well with the prediction from Ref.18. As both have
a maximum value between $Q^{2}=10$ and $100 GeV^{2}$ and then fall
to increase of $Q^{2}$ value. These results are in agreement to
the $\mathrm{k_{t}}$- factorization approach [27] only at low
$Q^{2}$ that it is continues to rise with increasing of $Q^{2}$
value. In Fig.7, we show the logarithmic $x$-derivatives of the
charm structure functions. In this figure, we can observe that
logarithmic $x$-derivatives of the charm structure functions are
constant at all $Q^{2}$ and $x$ values. In fact, it is more likely
that $x$- slope depends very weakly on $x$ and it is not depend on
$Q^{2}$. Therefore the Pomeron intercept for the charm structure
functions can be defined as
$\alpha_{P}=1+\lambda^{c}_{k}{\simeq}1.44$ and this is according
to exchange a hard Pomeron object. Consequently, we can observe
that $x$- slope of the charm structure
 function exponents ($\frac{{\partial}\lambda_{k}^{c}}{{\partial}ln\frac{1}{x}}$) does not depend on
 $x$. Therefore the logarithmic $x$-derivatives of the
charm structure functions are equal to the  charm- Pomeron
intercept as
$\lambda_{k}^{c}{\simeq}\frac{{\partial}lnF_{k}^{c}(x,Q^{2})}{{\partial}ln\frac{1}{x}}$.\\

%%%%%%%%%%%%%%%%%%%%%%%%%%%%%%%%%%%%%%%%%%%%%%%%%%%%%%%%%%%%%%%%%%%%%%
\centerline{\textbf{Summary}}

 In conclusion, our numerical results for the charm structure functions at low $x$
are obtained by applying the hard (Lipatov) pomeron behavior at
all $Q^{2}$ values in the NLO analysis. To confirm the method
 and results, the calculated values are compared with the $H1$  and ZEUS data  and other models on the
 charm
 structure functions, at small $x$. This behavior at low $x$ is consistent
with a effective power behavior for the charm structure functions
( $F^{c}_{k}(x,Q^{2})=A^{c}_{k}x^{-\lambda^{c}_{k}}$). We can
observed that the charm structure functions increase as usual, as
$x$ decreases. The form of the obtained distribution function for
the charm structure functions are similar to the predicted from
the proton paramerterization, and this is in agreement with the
increase observed by the $H1$ and ZEUS experiments. We shown that
the ratio $R^{c}=F_{L}^{c}/F_{2}^{c}$  for the charm structure
functions is constant at low- $x$ and all $Q^{2}$ values and its
independent of the input gluon distribution function. Results are
in agreement with those extracted in Refs.[19,27] within the all
uncertainty in the framework of perturbative QCD. We have
therefore obtained the charm exponents ($\lambda^{c}_{k}$) and
show that they are equal to the derivative of the charm structure
functions as
${{\partial}{\ln}F^{c}_{k}(x,Q^{2})}/{{\partial}{\ln}\frac{1}{x}}=\lambda^{c}_{k}$
 . This behavior
of the charm structure functions at low $x$ is consistent with a
power- law behavior and is in agreement with other models ( DL and
 color dipole models).\\

%%%%%%%%%%%%%%%%%%%%%%%%%%%%%%%%%%%%%%%%%%%%%%%%%%%%%%%%%%%%%%%%%%%%%%%%
\textbf{References}\\
1.Yu.L.Dokshitzer, Sov.Phys.JETP {\textbf{46}}, 641(1977);
G.Altarelli and G.Parisi, Nucl.Phys.B \textbf{126}, 298(1977);
V.N.Gribov and L.N.Lipatov,
Sov.J.Nucl.Phys. \textbf{15}, 438(1972).\\
 2. A.De Rujula \textit{et al}., phys.Rev.D \textbf{10}, 1649(1974);
 R.D.Ball and S.Forte, Phys.Lett.B \textbf{335}, 77(1994).\\
 3. A.Donnachie and P.V.Landshoff, Z.Phys.C \textbf{61},
139(1994); Phys.Lett.B \textbf{518}, 63(2001); Phys.Lett.B \textbf{533}, 277(2002); Phys.Lett.B \textbf{470}, 243(1999);
Phys.Lett.B \textbf{550}, 160(2002).\\
4. R.D.Ball and P.V.landshoff, J.Phys.G\textbf{26}, 672(2000).\\
5. P.V.landshoff, arXiv:hep-ph/0203084 (2002).\\
6. E.A.Kuraev, L.N.Lipatov and V.S.Fadin, Phys.Lett.B \textbf{60},
50(1975); Sov.Phys.JETP \textbf{44}, 433(1976); ibid. \textbf{45},
199(1977);
Ya.Ya.Balitsky and L.N.Lipatov, Sov.J.Nucl.Phys. \textbf{28}, 822(1978).\\
7. A.Vogt, arXiv:hep-ph:9601352v2(1996).\\
8. H.L.Lai and W.K.Tung, Z.Phys.C\textbf{74},463(1997).\\
9. A.Donnachie and P.V.Landshoff, Phys.Lett.B\textbf{470},243(1999).\\
10. N.Ya.Ivanov, Nucl.Phys.B\textbf{814}, 142(2009).\\
11. F.Carvalho, et.al., Phys.Rev.C\textbf{79}, 035211(2009).\\
12. S.J.Brodsky, P.Hoyer, C.Peterson and
N.Sakai,Phys.Lett.B\textbf{93}, 451(1980); S.J.Brodsky, C.Peterson
and N.Sakai, Phys.Rev.D\textbf{23}, 2745(1981).\\
13. A.V.Kotikov and G.Parente, Phys.Lett.B \textbf{379}, 195(1996).\\
14. K.Lipta, PoS(EPS-HEP)313,(2009).\\
15. C. Adloff et al. [H1 Collaboration], Z. Phys. C\textbf{72},
593 (1996); J. Breitweg et al. [ZEUS Collaboration], Phys. Lett.
B\textbf{407}, 402 (1997); C. Adloff et al. [H1 Collaboration],
Phys. Lett. B\textbf{528}, 199 (2002); S. Aid et. al., [H1
Collaboration], Z. Phys. C\textbf{72}, 539 (1996); J. Breitweg et.
al., [ZEUS Collaboration], Eur. Phys. J. C\textbf{12}, 35 (2000);
S. Chekanov et. al., [ZEUS Collaboration], Phys. Rev.
D\textbf{69}, 012004 (2004); Aktas et al. [H1 Collaboration], Eur.
Phys.J. C\textbf{45}, 23 (2006); F.D. Aaron et al. [H1
Collaboration],Phys.Lett.b\textbf{665}, 139(2008),
Eur.Phys.J.C\textbf{65},89(2010).\\
16. M.Gluk, E.Reya and A.Vogt, Z.Phys.C\textbf{67}, 433(1995); Eur.Phys.J.C\textbf{5}, 461(1998).\\
17. E.Laenen, S.Riemersma, J.Smith and W.L. van Neerven,
Nucl.Phys.B \textbf{392}, 162(1993).\\
18. A.~Y.~Illarionov, B.~A.~Kniehl and A.~V.~Kotikov, Phys.\
Lett.\  B {\bf 663}, 66 (2008).\\
19. S. Catani, M. Ciafaloni and F. Hautmann, Preprint
CERN-Th.6398/92, in Proceeding of the Workshop on Physics at HERA
(Hamburg, 1991), Vol. 2., p. 690; S. Catani and F. Hautmann, Nucl.
Phys. B \textbf{427}, 475(1994); S. Riemersma, J. Smith and W. L.
van Neerven, Phys. Lett. B \textbf{347}, 143(1995).\\
20. B.Rezaei and G.R.Boroun, JETP\textbf{112}, No.3,
381(2011).\\
21. H.L.Lai et. al., [CTEQ Collaboration], Eur.Phys.J.C\textbf{12}, 375(2000).\\
22. A.D.Martin, R.G.Roberts, W.J.Stirling and R.S. Thorn,
Eur.Phys.J.C\textbf{35}, 325(2004).\\
23. M.Froissart and A.Martin, Nuovo Cim.A\textbf{42}, 930(1965).\\
24. P.Desgrolard et.al., JHEP\textbf{02}, 029(2002).\\
25.M. Gluck, P. Jimenez-Delgado, E. Reya,
Eur.Phys.J.C\textbf{53},355(2008).\\
26. N.N.Nikolaev and V.R.Zoller, Phys.Lett. B\textbf{509},
283(2001).\\
27. A.~V.~Kotikov, A.~V.~Lipatov, G.~Parente and N.~P.~Zotov Eur.\
Phys.\ J.\  C {\bf 26}, 51 (2002).\\
28. V.~P.~Goncalves and M.~V.~T.~Machado, Phys.\ Rev.\ Lett.\ {\bf
91}, 202002 (2003).\\
29. N.Ya.Ivanov, Nucl.Phys.B\textbf{814} , 142 (2009).\\

%%%%%%%%%%%%%%%%%%%%%%%%%%%%%%%%%%%%%%%%%%%%%%%%%%%%%%%%%%%%%%%
 %%%%%%%%%%%%%%%%%%%%%%%%%%%%%%%%%%%%%%%%%%%%%%%%%%%%%%%%%%%%%%%%
\newpage{
\begin{figure}
\includegraphics[width=0.35\textwidth]{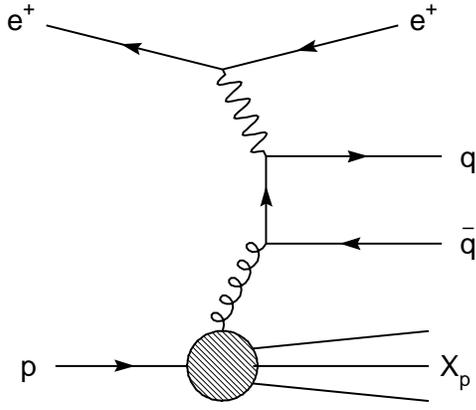}
\caption{The photon- gluon fusion} \label{Fig1}
\end{figure}
\begin{figure}
\includegraphics[width=0.5\textwidth]{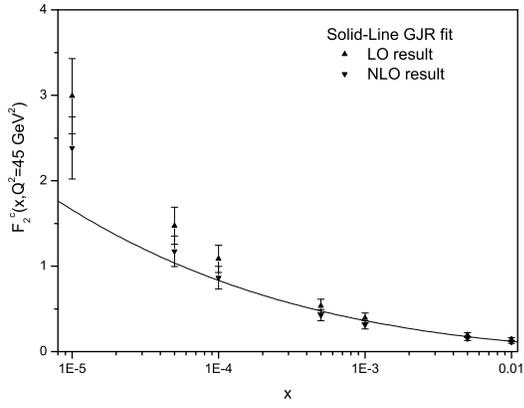}
\caption{$x$ dependence of the charm structure function
$F^{c}_{2}$ at $Q^{2}=45 GeV^{2}$. Plotted are the LO (Up
triangle) and NLO (Down triangle) our predictions that accompanied
to the errors due to the renormalization scales, as well as the
GJR parameterization [25](Solid line) results. }\label{Fig2}
\end{figure}
 \begin{figure}
\includegraphics[width=1\textwidth]{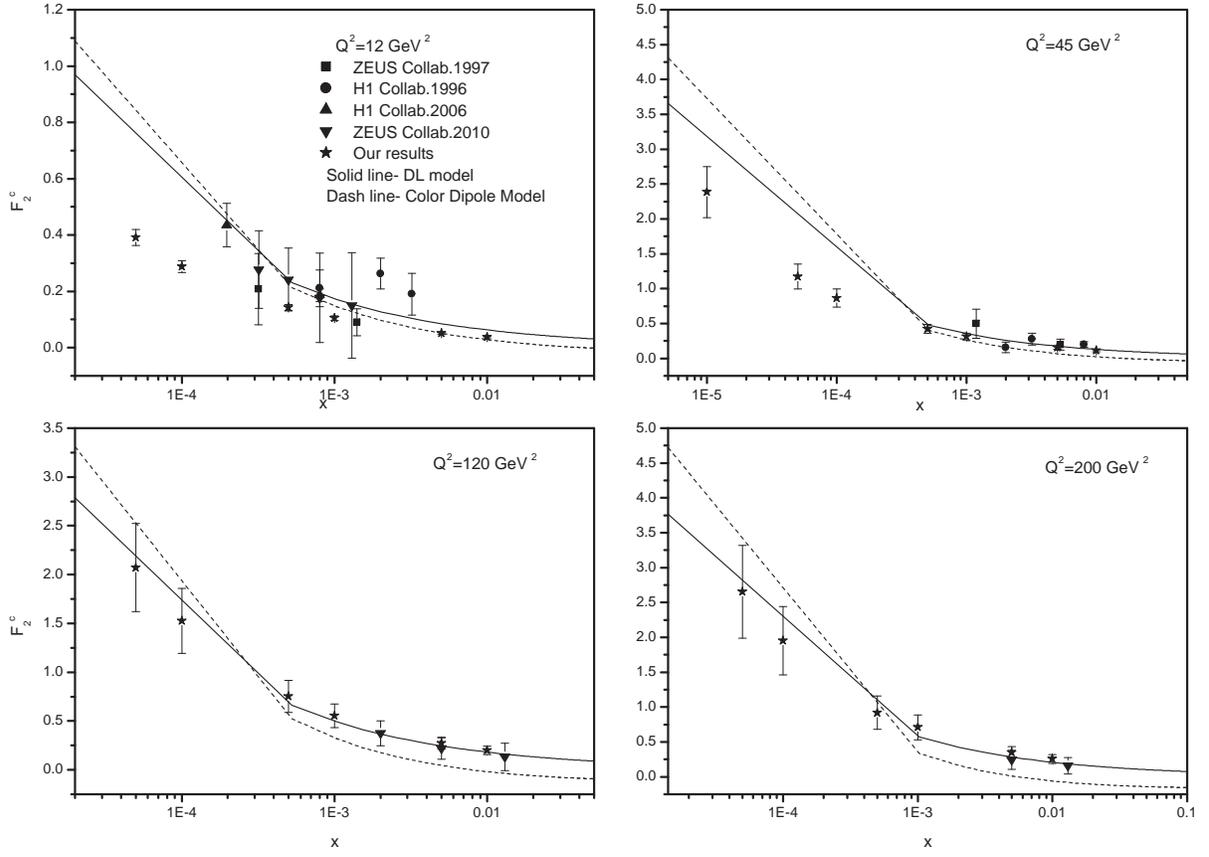}
\caption{The charm component of the structure function for
different values of
 $Q^{2}$ ($12 $, $45 $, $120
  $  and $200~ GeV^{2}$) as function of $x$. These results are the
  NLO predictions that accompanied to the theoretical uncertainty  related to the
renormalization scales. The solid and dash curves represents
$F^{c}_{2}$
  for DL [3-5] and color dipole [25] models.  Data are from H1 and ZEUS
  Collab. that accompanied to the total errors [15]. }\label{Fig3}
\end{figure}
\begin{figure}
\includegraphics[width=1\textwidth]{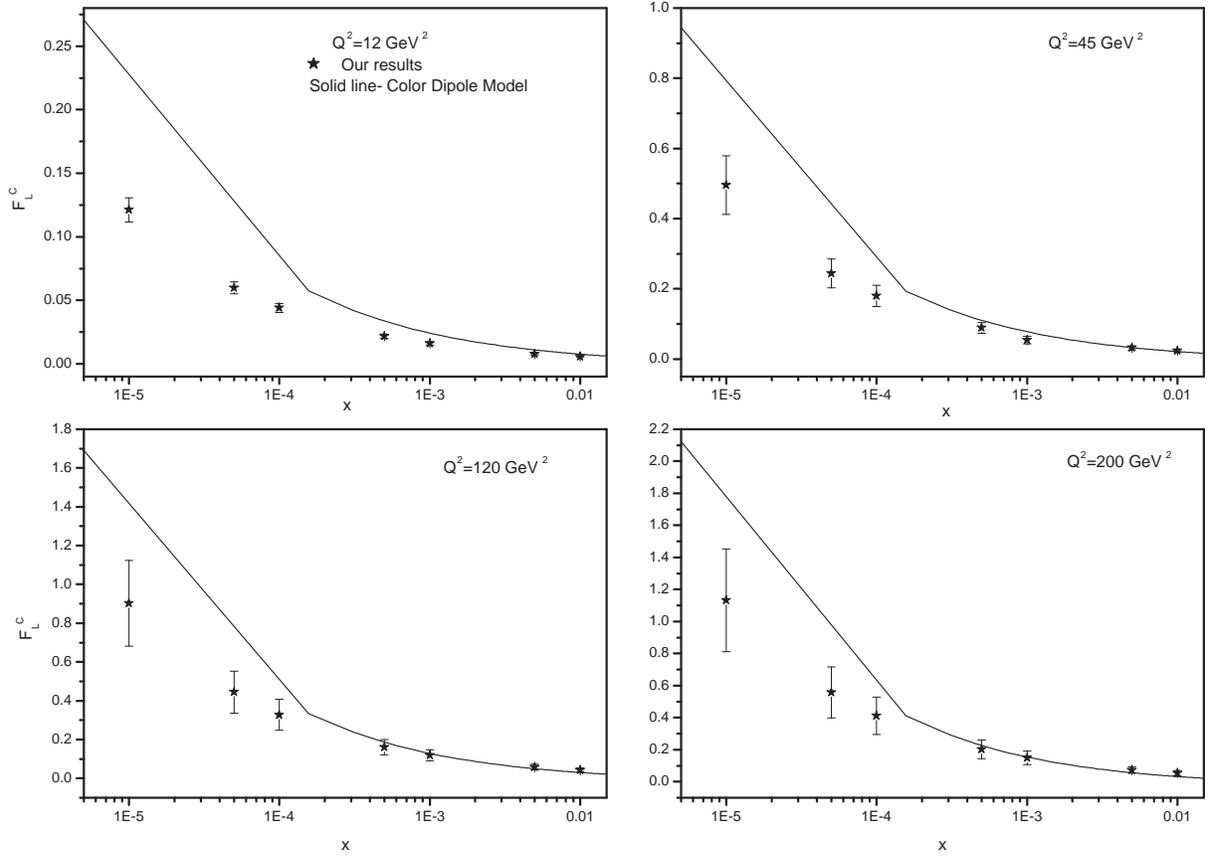}
\caption{The charm component of the  longitudinal structure
function at the
 $Q^{2}$ values $12$, 45, 120 and $200 ~GeV^{2}$ as function of
 $x$.These results are the
  NLO predictions that accompanied to the theoretical uncertainty  related to the
renormalization scales. The solid curves represents $F^{c}{_L}$
from the color dipole [25] model.  } \label{Fig4}
\end{figure}
%%%%%%%%%%%%%%%%%%%%%%%%%%%%%%%%%%%%%%%%%%%%%%%%%%%%%%%%%%%%%%%%%
\begin{figure}
\includegraphics[width=1\textwidth]{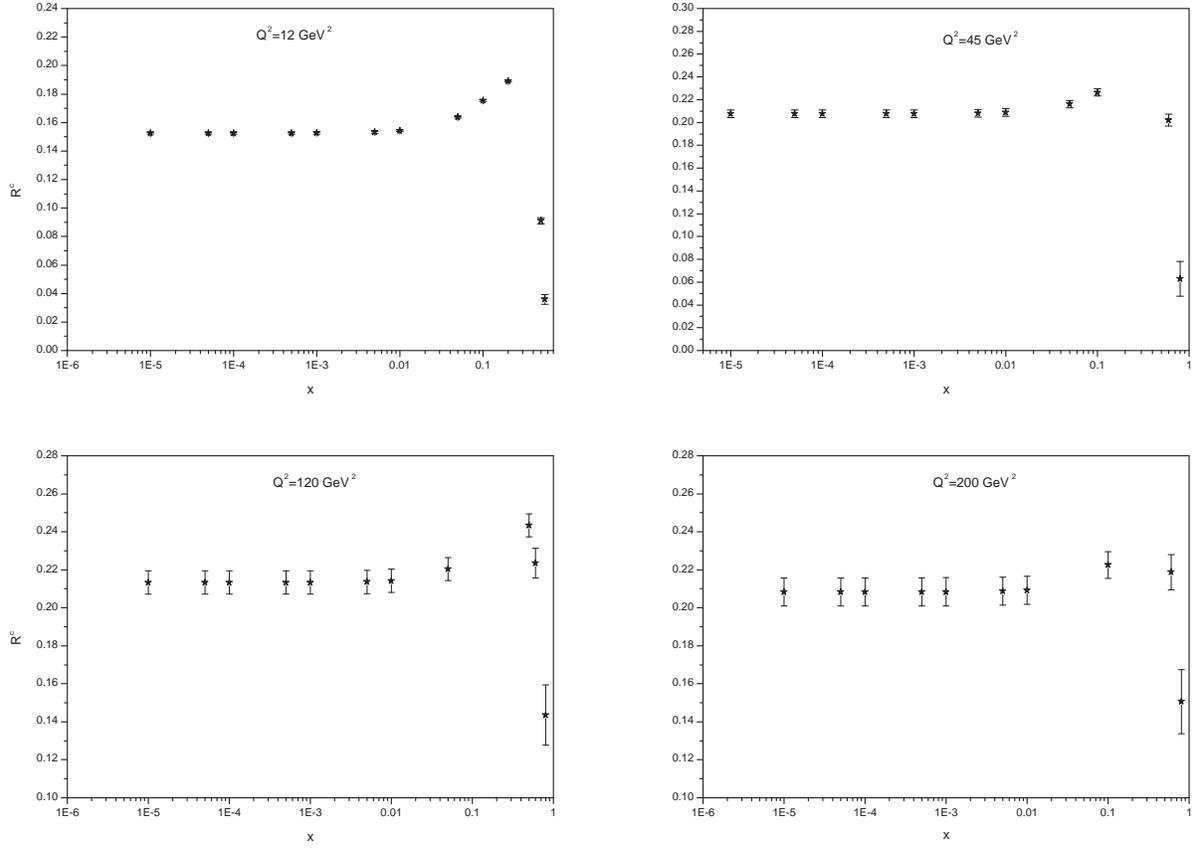}
\caption{The ratio $R^{c}=F_{L}^{c}/F_{2}^{c}$ as a function of
$x$ for different values of $Q^{2}$ in NLO analysis.} \label{Fig5}
\end{figure}
\begin{figure}
\includegraphics{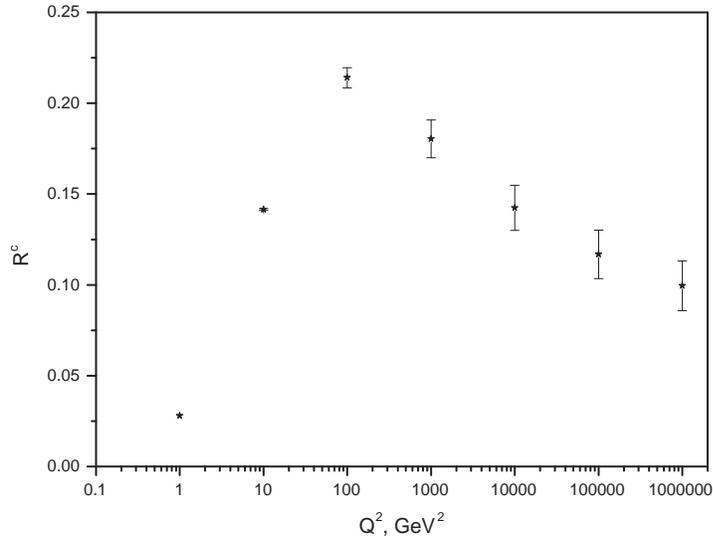}
\caption{The ratio $R^{c}=F_{L}^{c}/F_{2}^{c}$ as a function of
 $Q^{2}$ at $x=0.001$ in NLO analysis.} \label{Fig6}
\end{figure}
\begin{figure}
\includegraphics{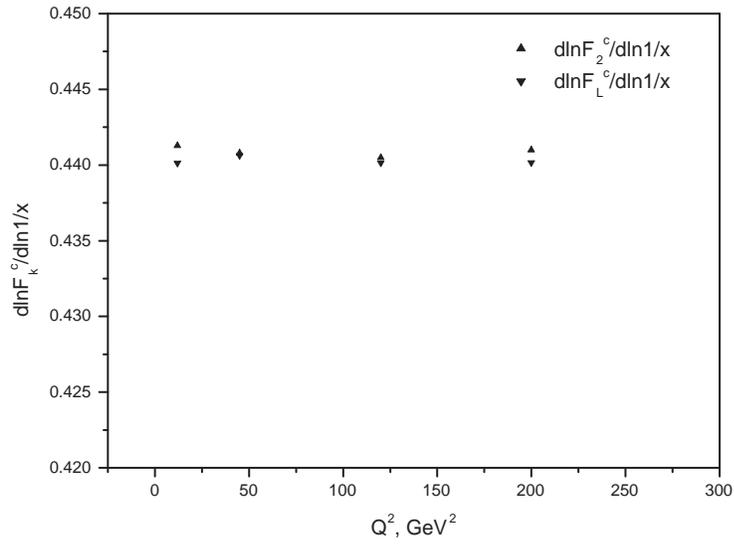}
\caption{Logarithmic $x$-derivatives of the charm structure
function $lnF_{k}^{c}(x,Q^{2})/ln\frac{1}{x}$ as function of
$Q^{2}$ values. } \label{Fig7}
\end{figure}}
%%%%%%%%%%%%%%%%%%%%%%%%%%%%%%%%%%%%%%%%%%%%%%%%%%%%%%

%%%%%%%%%%%%%%%%%%%%%%%%%%%%%%%%%%%%%%%%%%%%%%%%%%%%%%%%%%%%
\end{document}